\newcommand{\nn}{\nonumber}

\documentclass{jltp}
\usepackage[dvips]{graphicx}

\title{Stability of a Vortex in Spinor Bose-Einstein Condensate}

\author{Tomoya Isoshima}

\address{Department of Physics, Okayama University, Okayama, 700-8530, Japan}

\runninghead{Tomoya Isoshima}{Stability of a Vortex in Spinor Bose-Einstein Condensate}

\begin{document}

\maketitle

\begin{abstract}
We propose a method to create a vortex in BEC utilizing the spin degree
of freedom.
We consider the optical plug
at the center of the vortex, which makes the vortex-creation
process stable.
We also investigate the instability of the halfway state to complete vortex
state without the optical plug.

\end{abstract}

\section{INTRODUCTION}

The vortex formation in
Bose Einstein condensate of atom gases has been
achieved in several groups~\cite{matthews,madison,mit,anderson,inouye,hodby,jila-t}
in variety of methods.
The first method is called phase imprinting~\cite{matthews}
and an another method rotate the potential which confine
the condensate.~\cite{madison}
We proposed another method in Ref.\ \onlinecite{persist},
which utilized the turning magnetic field.
This process is made stable using pinning potential with dipole force
at the center of the condensate.
This makes the condensate ring shaped
and the refined properties of vortices around the core is hidden.
So it is more desirable to form the vortex without the pinning potential.
In this paper, we compare the stability of condensate
with and without the pinning potential.
We analyze the excitation spectrum to obtain
the insight to the stability of the process.

The next section introduces Bogoliubov equations.
Section 3 is devoted to the explanation of method to create the vortex state.
In Section 4, we discuss the instability of vortex creation process.
Section 5 is devoted to summary.

\section{BOGOLIUBOV EQUATIONS}

In order to treat an interacting Boson system, we use the formulation
based on Bogoliubov theory, which is extended to treat the
spin degrees of freedom.
The wavefunction of condensate is written with
a set of 3 wavefunctions $\phi_i({\bf r}) (i = 0, \pm 1)$.
These are given by the Gross-Pitaevskii (GP) equation
\begin{eqnarray}
\biggr[
( - \frac{\hbar^2 \nabla^2}{2m} + V(r) + g_n \sum_k |\phi_k|^2 )\delta_{ij}
    -{\mathcal B}_{ij}
&&\nn\\
    + g_s \sum_\alpha \sum_{kl}
    (F_\alpha)_{ij} (F_\alpha)_{kl} \phi_k^\ast \phi_l
\biggl] \phi_j &=& 0,
\label{eq:gp}
\end{eqnarray}
where $\mu$ is the chemical potential.
The coefficient $g_n$ and $g_s$ are interaction parameters.
The numbers ${\mathcal B}_{ij}$ are matrix elements of ${\mathcal B}$-matrix
which shows the magnetic field. The form of this matrix is given below.
The subscripts are $\alpha = (x,y,z)$ and $i,j,k,l = (0, \pm 1)$.
$V(r)$ is the spin-independent potential which is given in Eqs.\ (\ref{eq:pot1}) and (\ref{eq:pot2}).

The excitation levels $\varepsilon_q$ and
corresponding wavefunctions $u_q, v_q$ are given by Bogoliubov equations
\begin{eqnarray}
\sum_j \{ A_{ij}        u_{q}({\bf r},j) - B_{ij}        v_{q} ({\bf r}, j) \}
             &=& \varepsilon_q u_q({\bf r}, i),
\label{eq:bog1}
\\
\sum_j \{ B_{ij}^{\ast} u_{q}({\bf r},j) - A_{ij}^{\ast} v_{q} ({\bf r}, j) \}
             &=& \varepsilon_q v_q({\bf r}, i)
\label{eq:bog2}
\end{eqnarray}
where
\begin{eqnarray}
A_{ij} &=& \left(
   -\frac{\hbar^2 \nabla^2}{2m} - \mu + V({\bf r}) + g_n \sum_k |\phi_k|^2
\right) \delta_{ij}
-{\mathcal B}_{ij}
\nn\\&&
  + g_n \phi_i^{\ast} \phi_j 
  + g_s \sum_{\alpha} [(F_\alpha)_{ij}(F_\alpha)_{kl} \phi_k^{\ast}\phi_l
  + (F_\alpha)_{il}(F_\alpha)_{kj} \phi_k^{\ast}\phi_l],
\label{eq:bog3}
\\
B_{ij} &=& g_n \phi_i \phi_j + g_s \sum_\alpha \sum_{kl}
    (F_{\alpha})_{ik} \phi_k (F_\alpha)_{jl}\phi_l.
\label{eq:bog4}
\end{eqnarray}
Most of $\varepsilon_q$ are positive.
When the lowest $\varepsilon_q$ is negative,
it shows the local instability of condensate.

\section{PROCESS TOWARD THE VORTEX}

We consider a two dimensional system of BEC
with rotational symmetry around the $z$ axis.
We introduce the cylindrical coordinates ${\bf r} = (r, \theta, z)$.
Suppose that an Ioffe-Pritchard field
\begin{equation}
  {\bf B} = (B_{\perp}\cos \theta, -B_{\perp} \sin \theta, B_z)
\label{eq:jiba}
\end{equation}
is applied to the system.
The coefficients $B_{\perp}$ and $B_z$ are amplitude of
magnetic field measured by Zeeman energy.
This magnetic field is rewritten in matrix form as
\begin{equation}
   {\mathcal B} = \left(
      \begin{array}{ccc}
         B_z &  B_{\perp} \frac{e^{i \theta}}{\sqrt{2}}   &  0 \\
         B_{\perp} \frac{e^{- i \theta}}{\sqrt{2}} & 0 &  B_{\perp} \frac{e^{i \theta}}{\sqrt{2}} \\
         0   &  B_{\perp} \frac{e^{-i \theta}}{\sqrt{2}} & -B_z
    \end{array}
   \right).
\label{eq:b-mat}
\end{equation}
The eigenvectors of ${\mathcal B}$ show the arrangement of spin components
when the field is strong enough.
The eigenvalues of ${\mathcal B}$ are $0$ and $\pm B$ where
$B = \sqrt{B_\perp^2 + B_z^2}$.
The eigenvectors corresponding to $\pm B$ are 
\begin{eqnarray}
   \left( \begin{array}{c}
      \phi_{+1} \\ \phi_{0} \\ \phi_{-1} \\
   \end{array}
   \right) = 
   \frac{1}{2B} \left(
     \begin{array}{c}
        (B \pm B_z) e^{i \theta} \\
        \pm \sqrt{2} B_{\perp} \\
        (B \mp B_z) e^{- i \theta}
     \end{array}
  \right).
\label{eq:b-eignvec}
\end{eqnarray}
The winding number
(coefficient of $\theta$ in exponential)
becomes $1, 0,$ and $-1$ for $\phi_1, \phi_0$ and $\phi_{-1}$
respectively.
When the coefficients $B_{\perp}$ and $B_z$ vary as
\begin{eqnarray}
B_z &=& B_{z0} \cos [\pi ( 1 - t / T) ]               \label{eq:bz}
\\
B_\perp &=& B_{\perp}^{\prime} r \sin[\pi ( 1 - t / T) ]     \label{eq:bperp}
\end{eqnarray}
where $t$ $(0 \le t \le T)$ is time,
the spin of condensate shifts following Eq.\ (\ref{eq:b-eignvec}).
If the process begins with condensate of -1 component without vortex,
the initial arrangement of condensate is Eq.\ (\ref{eq:b-eignvec})
multiplied by $e^{i\theta}$.
The coefficient of $\theta$ in exponential of $\phi_{-1}$ is zero and
the condensate takes the form
\begin{equation}
   \left( \begin{array}{c}
      \phi_{+1} \\ \phi_{0} \\ \phi_{-1}
   \end{array}
   \right) \propto 
   \left( \begin{array}{c}
        (B + B_z) e^{2i \theta} \\
        \sqrt{2} B_{\perp} e^{i \theta} \\
        (B - B_z)
   \end{array}
   \right).
\label{eq:phiswind}
\end{equation}
This process ends with condensate of $+1$ component with
winding 2 at $t=T$.
This means that the vortex is nucleated.

The explanation above treat only the magnetic field.
We can take into account other terms with GP equation Eq.\ (\ref{eq:gp}).
This does not change the arrangement of winding numbers
as long as the condition of cylindrical symmetry is kept.

When the condensate follows Eq. (\ref{eq:phiswind}),
the winding numbers of excitation levels are written as
$q_{\theta} + 2, q_{\theta}+1, q_{\theta}$ for $u_{+1}, u_{0}, u_{-1}$
and
$q_{\theta} - 2, q_{\theta}-1, q_{\theta}$ for $v_{+1}, v_{0}, v_{-1}$
respectively.
When the pinning potential does not exist,
the modes with $q_{\theta} = \pm 1 \mbox{ and } \pm 2$ can localize at the core
(and have negative energy)
because they have winding number 0 in $u_i \mbox{ or } v_i$.

\section{ANALYSIS OF STABILITY}

The stability of the condensate can be evaluated
with sign of the lowest excitation level, which is given by
Bogoliubov equations Eqs.\ (\ref{eq:bog1}) and (\ref{eq:bog2}).
We decide the distribution of condensate with GP equation Eq.\ (\ref{eq:gp})
at each time $t$, and the excitation levels with
Eqs.\ (\ref{eq:bog1}) and (\ref{eq:bog2}).
The whole condensate is confined with spin-independent
harmonic potential, and the magnetic field which is described in
Eqs.\ (\ref{eq:bz}) and (\ref{eq:bperp}) handle the spin of condensate.
The parameters are
$m = 1.44 \times 10^{-25}$ kg, 
$a_0 = 5.5 \times 10^{-9}$ m,
$g_n = \frac{4 \pi \hbar^2}{m}\frac{2 a_0 + a_2}{3}$,
$g_s = \frac{4 \pi \hbar^2}{m}\frac{a_2 - a_0}{3}$.
The scattering length $a_2$ is 
$0.75 a_0$ or $1.375 a_0$, which means $g_s=-0.1 g_n$ and
$g_s = +0.1 g_n$ respectively.

The case when the pinning potential at the center of the system
is introduced is discussed in next subsection,
and the case without the pinning potential is discussed in 
subsection \ref{subsec:nopin}
The time $T$ is supposed to be enough longer than the Larmor frequency,
namely the adiabatic condition is satisfied.
When the time $T$ is too short, the spin of atoms will not
have enough time to follow the perpendicular magnetic field.
Whether the vortex nucleates or not is uncertain in this case.

\subsection{WITH PINNING POTENTIAL}

The whole condensate is confined with spin-independent
harmonic potential radially, and
the pinning potential at the $z$-axis of the system is also given.
This stabilizes the vortex creation process.
These two spin-independent potentials are written as
\begin{equation}
   V(r) = \frac{m (2 \pi \nu)^2}{2} r^2  + U \exp \left( \frac{ - r^2}{2 r_0^2 } \right)
\label{eq:pot1}
\end{equation}
where $\nu = 200 {\rm Hz}, U = 2 \times 10^3 {\rm J}$
and $r_0 = 1 \mu m$.

\begin{figure}
\begin{center}
\begin{tabular}{cc}
    \includegraphics[width=2.1in]{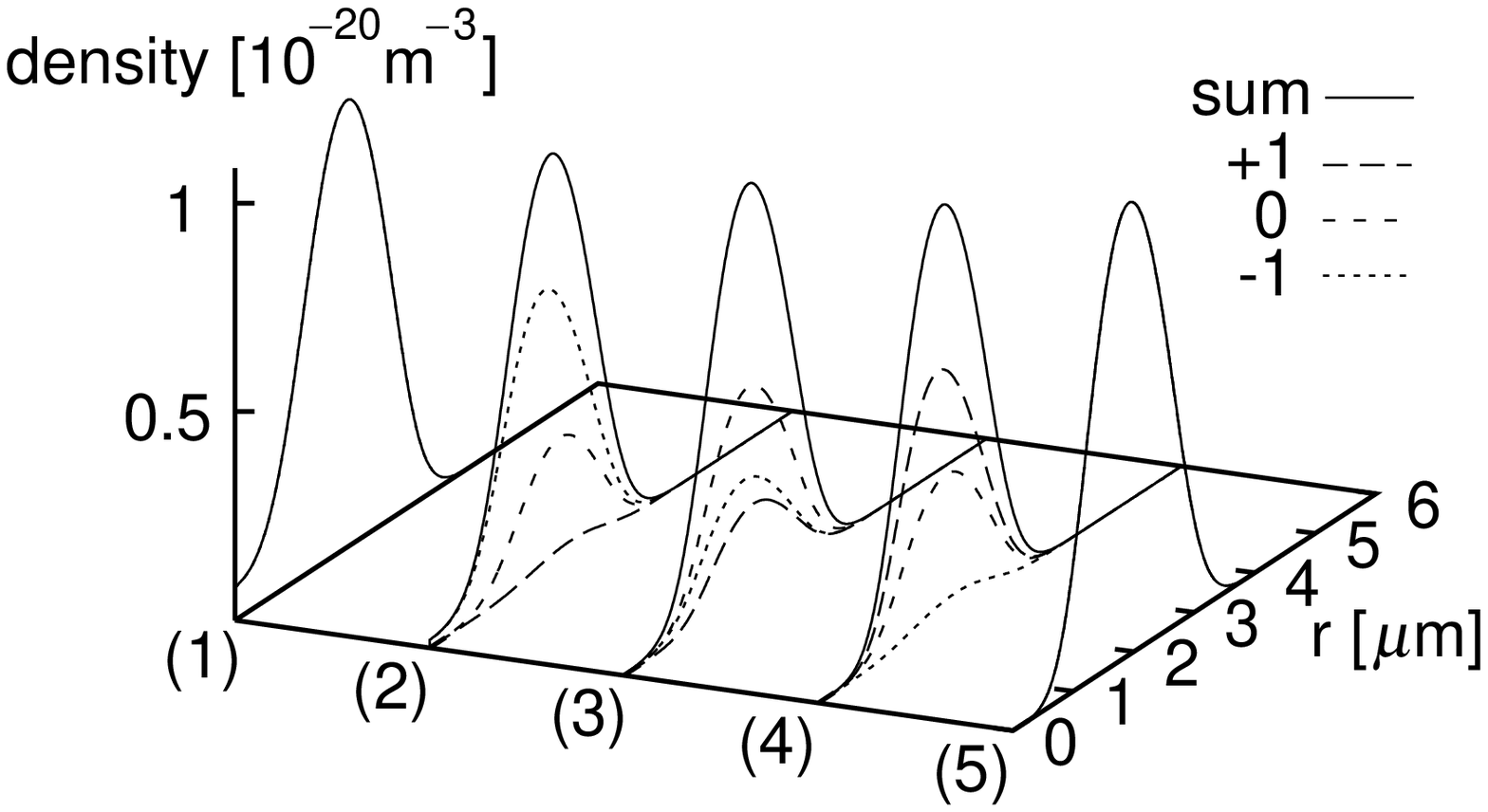}
            & \includegraphics[width=2.1in]{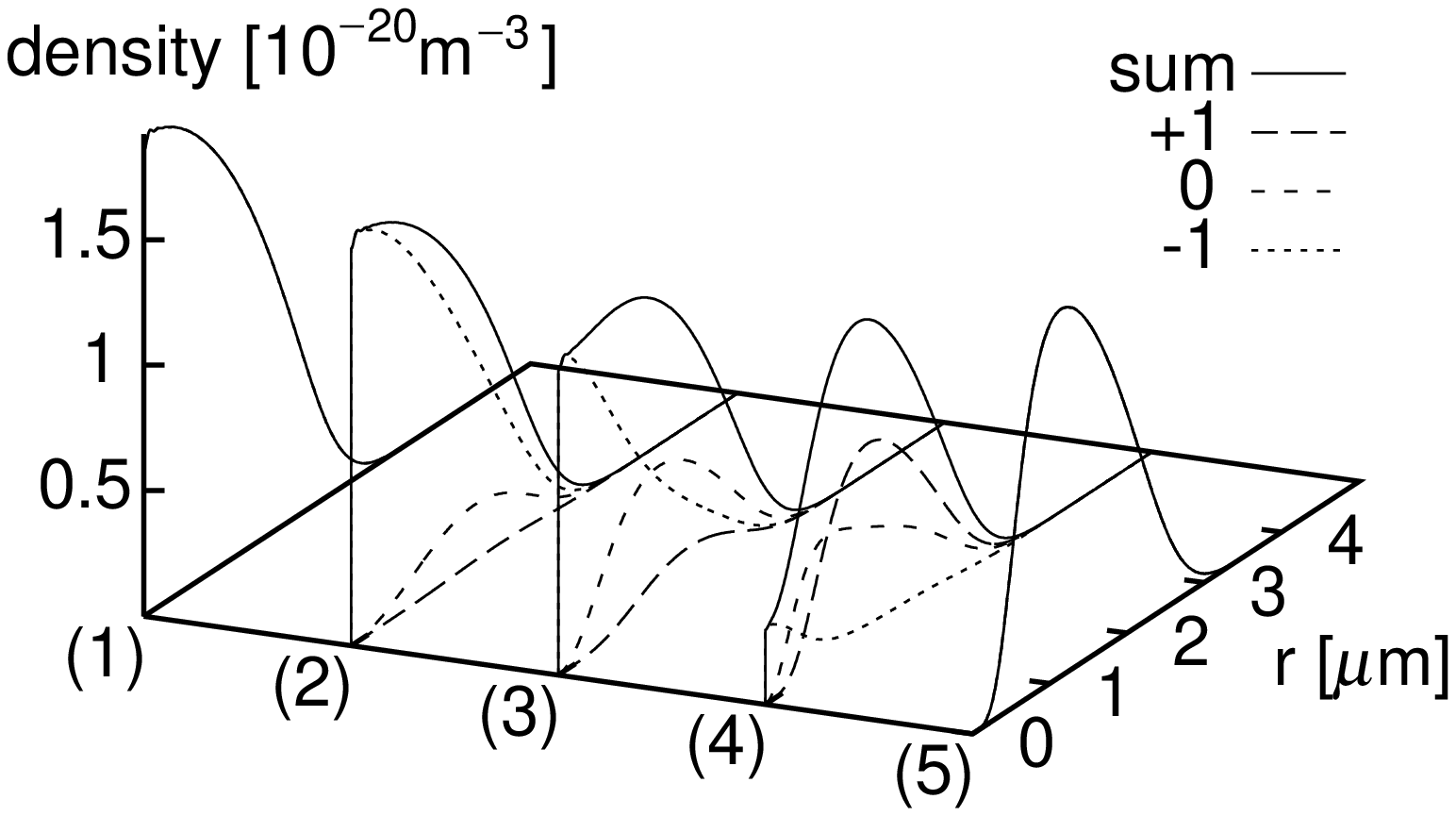}
\\
    \raisebox{3mm}[1mm][-1mm]{(a)} & \raisebox{3mm}[1mm][-1mm]{(c)}
\\
    \hspace{-0.2in}\includegraphics[width=2.0in]{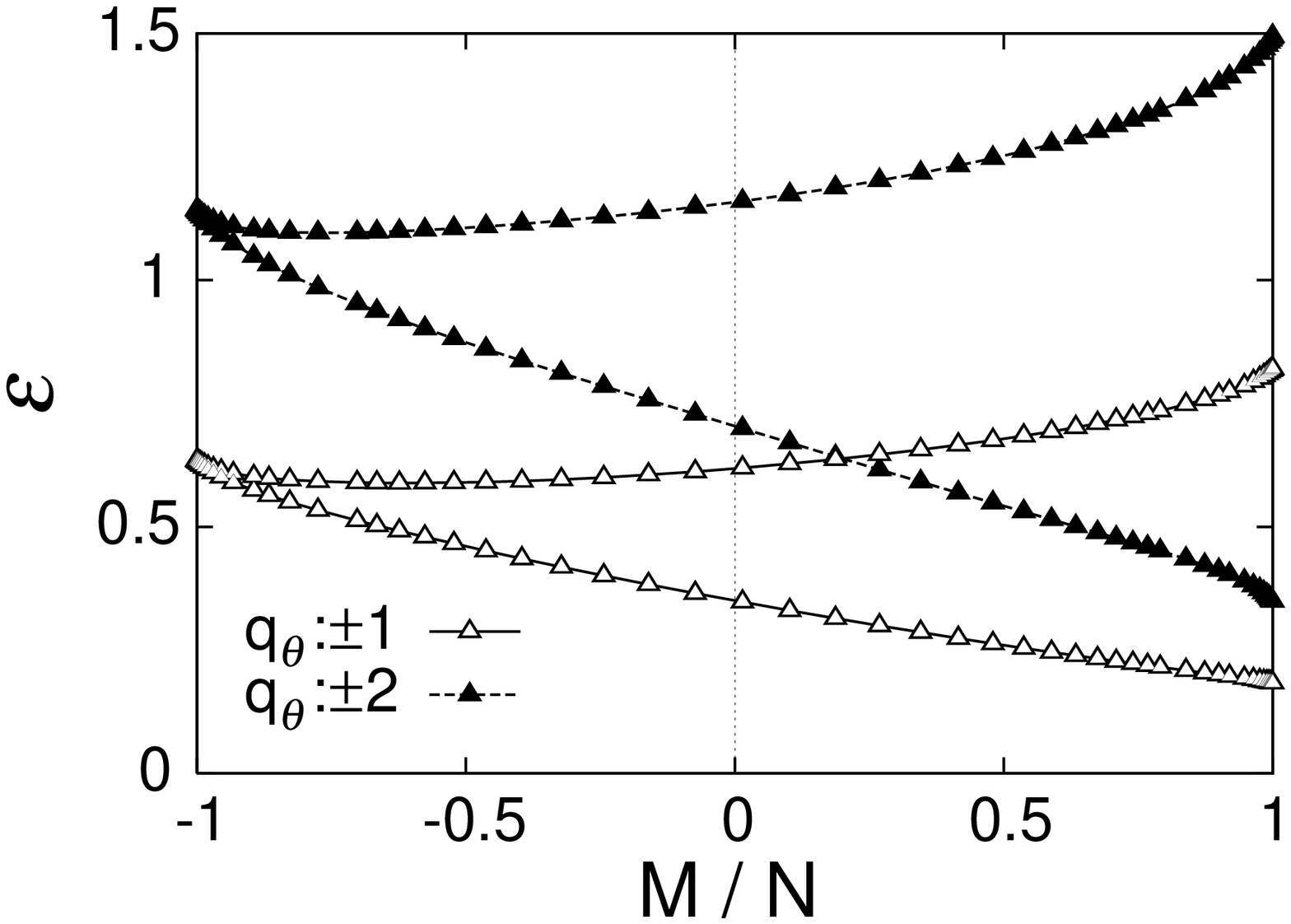}
            & \hspace{-0.2in}\includegraphics[width=2.0in]{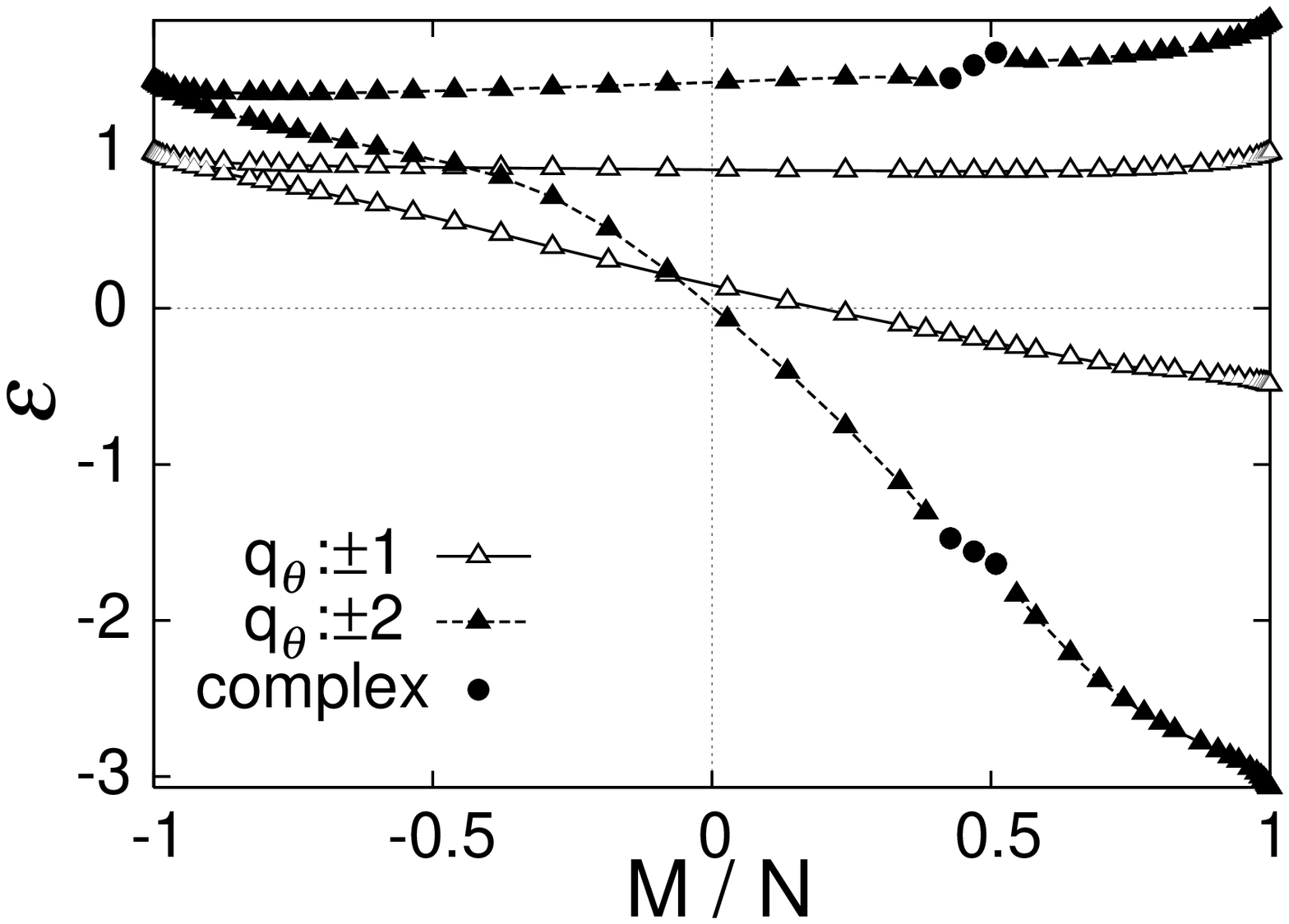}
\\
    \raisebox{2mm}[0.5mm]{(b)} & \raisebox{2mm}[0.5mm]{(d)}
\\
     & \hspace{-0.2in}\includegraphics[width=2.0in]{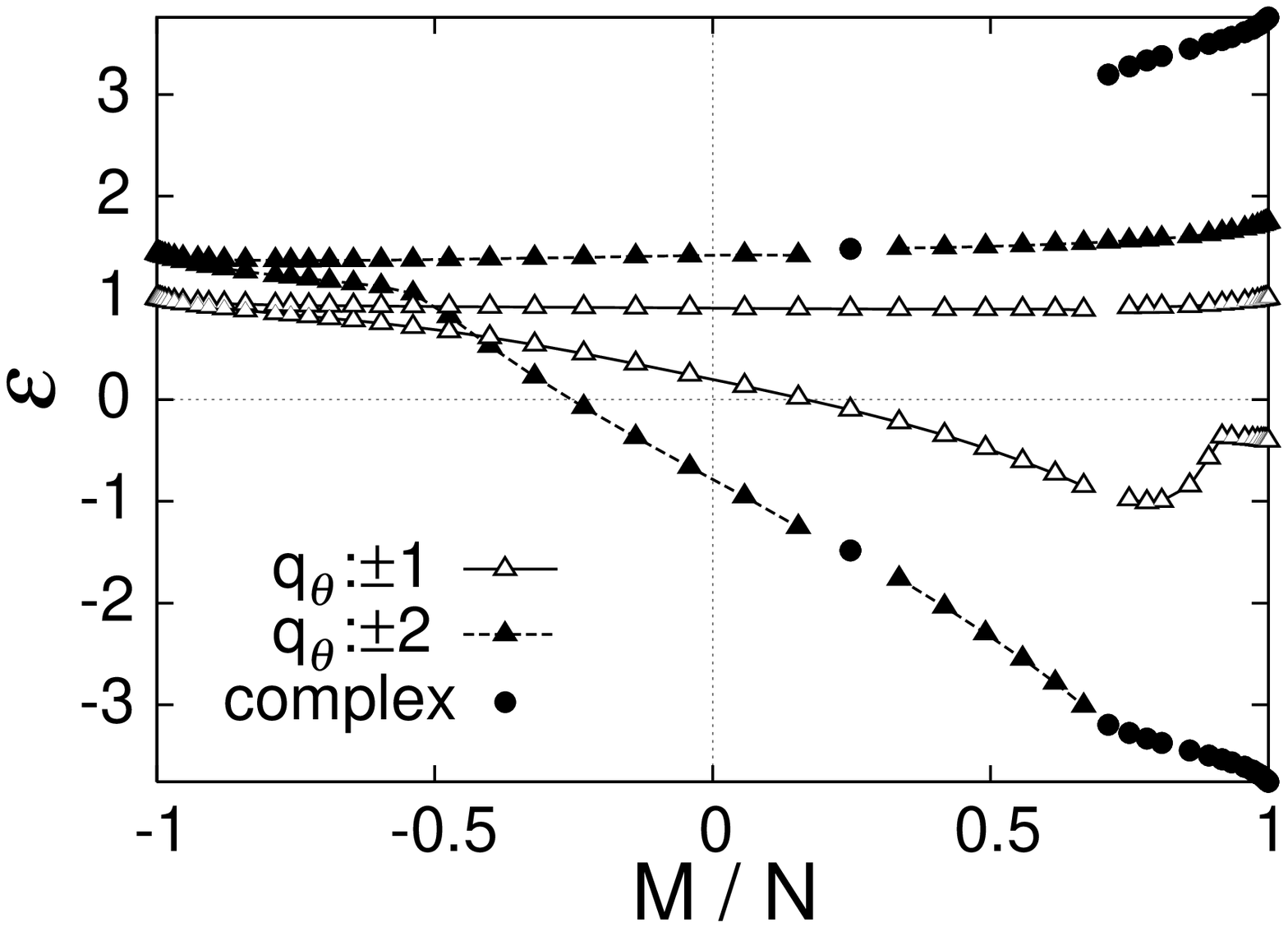}
\\
     & \raisebox{2mm}[1mm]{(e)}
\end{tabular}
\end{center}

\vspace{-5mm}
\caption{
In (a) and (c), the solid line shows
the density profile of condensate $\sum |\phi_i|^2$
and the various dashed lines show the density of
three spin components $|\phi_i|^2$ $(i = 0, \pm 1)$.
In (b), (d) and (e), the lowest eigenenergy $\varepsilon$
at winding number index $q_\theta = \pm 1, \pm 2$ are plotted.
The unit of $\varepsilon$ is $200 h J$, which is the
trap unit when the pinning potential is not exist.
The horizontal axis show
magnetization along z-axis which varies from -1 to +1 during the process.
(a) and (b) show shift of density distribution of the condensate
and the lowest eigenenergies when the pinning potential exists.
The eigenenergies never become negative.
(c), (d) and (e) show those when the pinning potential does not exist.
Some of eigenenergies become negative and even complex
as the magnetization increases.
In (a), (b), (c) and (d) the interaction parameter $g_s = -0.1$, which
means ferromagnetic interaction. $a_2$ is $0.75 a_0$.
In (e), the interaction parameter $g_s = +0.1$, which
means antiferromagnetic interaction and $a_2$ is $1.375 a_0$.
The density distributions (a) and (b) hardly change when
the sign of interaction parameter $g_s$ is turned to antiferromagnetic.
}

\label{fig:both}
\end{figure}

The magnetic field which is given by Eq.\ (\ref{eq:bz}) and (\ref{eq:bperp})
are applied.
The wavefunctions of the condensate and the excitation spectrum
are calculated at each $t$,
at each pairs of magnetic field coefficient $B_r$ and $B_{\perp}^{\prime}$.
Fig.\ \ref{fig:both}(a) shows the density distribution of the condensate
at several $t$.
Fig.\ \ref{fig:both}(b) shows the lowest excitation levels at $q_{\theta} = \pm 1, \pm 2$.
The change of total magnetization $M$
along the $z$-axis accompanies the change of $t$,
and the horizontal axis of Fig.\ \ref{fig:both}(b) shows $M/N$,
which is $\int (|\phi_{1}|^2 - |\phi_{-1}|^2) d{\bf r}/ \sum_i \int |\phi_i|^2 d{\bf r}$.
It is found that the lowest $\varepsilon$ never becomes negative.
Because the negative excitation level does not appear,
the condensate stays locally stable.

\subsection{WITHOUT PINNING POTENTIAL} \label{subsec:nopin}

In this subsection,
we consider the case when the optical pinning does not exist.
The spin-independent trapping potential written in Eqs.\ (\ref{eq:gp}) 
and (\ref{eq:bog3}) is
\begin{equation}
   V(r) = \frac{ m (2 \pi \nu)^2}{2} r^2 ,
\label{eq:pot2}
\end{equation}
and the time-dependent magnetic field which is given
in Eqs.\ (\ref{eq:bz}) and (\ref{eq:bperp}) is also applied.
The halfway states are obtained by the static form
of Gross-Pitaevskii equation.
Figure \ref{fig:both}(c) shows some of the density distributions of
condensate.
Figure \ref{fig:both}(d) and  \ref{fig:both}(e) show
the excitation levels obtained through
Bogoliubov equations Eqs.\ (\ref{eq:bog1}) and (\ref{eq:bog2}).
The filled dot shows that there is a pair of eigenstate with complex eigenenergy.
The real parts of them are plotted.
The complex eigenenergy appears where the
$q_{\theta}=2$ mode and the $q_{\theta}=-2$ mode crossed,
which result is consistent with Ref.\ \onlinecite{complex}.
This fact means the spiraling out or the splitting of the vortex.

\section{CONCLUSION}

In this paper, we analyzed the stability of vortex creation process.
The distributions of condensate and excitation levels are derived
using Bogoliubov equations.
When the pinning potential is absent,
negative and complex eigenenergies are found and
the vortex forming become unstable halfway.

\end{document}